\documentclass[12pt]{article}
\usepackage{spconf}
\usepackage[utf8]{inputenc}
\usepackage{amsmath}
\usepackage{amsfonts}
\usepackage{amssymb}
\usepackage{booktabs}
\usepackage{cite}
\usepackage{color}
\usepackage[pdftex,final]{graphicx}
\usepackage{hyperref}
\usepackage{subfigure}
\usepackage{url}
\usepackage{epstopdf}
\usepackage{multirow}
\usepackage{adjustbox}
\newcommand{\ttcaps}[1]{{\small {\tt{#1}}}}

\newcommand{\on}{{\sc on~}}
\newcommand{\off}{{\sc off~}}

\usepackage[symbol]{footmisc}

\title{WIDEFT: A Corpus of Radio Frequency Signals for\\Wireless Device Fingerprint Research}

\name{
\begin{minipage}{\linewidth}
\centering  Abu Bucker Siddik$^1\!\!$, ~Dawson Drake$^{1,2}\!\!\!\!\!\!$, ~~~~Thomas Wilkinson$^{1,2}\!\!\!\!\!\!\!~$, ~~~Phillip L.~De Leon${}^{1}\!\!$, \\  Steven Sandoval$^{1}\!\!$, ~{\normalfont and} Margaret Campos$^2$
\end{minipage} \vspace*{-2mm}
}

\address{New Mexico State University\\
	$^1$Klipsch School of Electrical and Computer Engineering\\
	$^2$Physical Science Laboratory\\
	Las Cruces, New Mexico, U.S.A.\\	
	{\tt \{siddik,dvd82000,wthomasw,pdeleon,spsandov,mcampos\}@nmsu.edu}}

\begin{document}
\ninept
\maketitle

\begin{abstract}
Wireless network security may be improved by identifying networked devices via traits that are tied to hardware differences, typically related to unique variations introduced in the manufacturing process. One way these variations manifest is through unique transient events when a radio transmitter is activated or deactivated. Features extracted from these signal bursts have in some cases, shown to provide a unique ``fingerprint'' for a wireless device. However, only recently have researchers made such data available for research and comparison. Herein, we describe a publicly-available corpus of radio frequency signals that can be used for wireless device fingerprint research. The WIDEFT corpus contains signal bursts from 138 unique devices (100 bursts per device), including Bluetooth- and WiFi-enabled devices, from 79 unique models. Additionally, to demonstrate the utility of the WIDEFT corpus, we provide four baseline evaluations using a minimal subset of previously-proposed features and a simple ensemble classifier.
\end{abstract}


\section{Introduction}
\label{sec:intro}
Providing security in a wireless network environment is a challenging problem because access to network resources is possible without being physically connected. One way to enhance security is through unique identifiers associated to particular network devices without considering easily forgeable identifiers such as MAC addresses and user credentials \cite{zhang2003}. Prior research has demonstrated the ability to identify a device via physical, hardware-level traits due to variations/imperfections in the circuitry associated with manufacturing process \cite{banerjee2011}. One way these variations/imperfections manifest is through unique transient events when an wireless transmitter is activated or deactivated. Features extracted from these transient events may provide a unique ``fingerprint'' for a wireless device that is similar to a biometric for human identification \cite{hall2003detection}. Thus, device fingerprinting can offer an additional defensive layer of security beyond conventional (software) protocols for authentication \cite{danev2012physical, sieka2006using}. 

More specifically, hardware-based device-level fingerprinting offers additional security and safeguards against attacks such as device cloning, message replay, and spoofing attacks \cite{rasmussen2007}. On the other hand, fingerprinting may also be used offensively to gain information about network operations or to gain information about specific network users \cite{rasmussen2007}. A skilled attacker, for example, may be able to exploit a device fingerprint to violate sender anonymity, in order to associate a transmission to a specific sender. For example, FM transmitters have been shown in many cases, to have relatively short transient characteristics directly following when the transmitter is activated or ``keyed'' \cite{ellis2001characteristics}. Potentially, these transient characteristics are similar enough that the device model may be identifiable, and in the best case are unique enough to allow identification of an individual device. Recently, the widespread availability/use of software defined radios (SDRs) has lowered the barrier to adversarial cloning and spoofing \cite{vo2016fingerprinting}.  One potential defense against these attacks could be device fingerprinting. 

In this work, we describe a publicly-available data corpus for WIreless DEvice FingerprinTing (WIDEFT) research. In addition, we provide details of a feature vector and classifier in order to provide baseline evaluation results using the WIDEFT corpus. In Section \ref{sec:wideftcorpus}, we provide some background information on an existing corpus and our motivation for WIDEFT. This section carefully describes the devices for which we collect signal bursts from, the data collection environment, and our procedure for collecting the signal data. In Section \ref{sec:postprocessingorganization}, we describe the signal postprocessing which includes segmentation and the organization of the resulting files. In Section \ref{sec:Classification}, we provide results from four baseline evaluations using a minimal subset of previously-proposed features and a simple ensemble classifier. In Section \ref{sec:discussion}, we provide a short discussion on the WIDEFT corpus and the evaluation results and finally in Section \ref{sec:conclusion}, we conclude the article.

\section{WIDEFT Corpus}
\label{sec:wideftcorpus}
To the best of our knowledge there exists at least one recent paper which describes a publicly-available data corpus for wireless device fingerprinting. In \cite{uzundurukan2020}, the authors focus their data collection on Bluetooth devices which consisted of 86 individual smartphones from a set of 27 different models. The data consists of wireless signals acquired in a lab environment using a network analyzer at different sample rates. The signals are subsequently processed by transformation to the analytic signal via the Hilbert transform (HT), digitally downconverted, and segmented to preserve the \on transient and a portion of the steady-state which immediately follows. The authors also present experimental results for classification of the devices using features extracted from the signal transient \cite{uzundurukan2020}.

Unlike the data set in \cite{uzundurukan2020}, WIDEFT includes not only Bluetooth devices (e.g.~wireless ear buds, keyboards, mice) but also WiFi devices (e.g.~wireless routers, wireless media servers) and other 900 MHz devices (e.g.~wireless clickers). In some devices, such as laptops and smartphones, both Bluetooth and WiFi signals were acquired. Unlike \cite{uzundurukan2020}, we collect a complete radio burst consisting of the \on transient, steady-state portion, and  \off transient as all three portions may offer discriminating features to allow for improved fingerprinting. Our motivation for collecting a wide variety of devices is to allow research into the uniqueness (or lack thereof) of device fingerprints. This corpus of wireless bursts is made publicly-available to researchers on Zenodo (DOI 10.5281/zenodo.4110980) \cite{zenodo2020}.

\subsection{Device Types and Distribution}
\label{ssec:device}
The WIDEFT data corpus is broken up into three wireless types: Bluetooth, WiFi, and Other. The data is collected from 138 devices spanning 79 models\footnote{Some models have both Bluetooth and WiFi wireless types hence why the number of unique Bluetooth models and unique WiFi models do not sum to the total number of unique models.} and is summarized in Table \ref{tbl:WirelessTypeDistribution}. Each of the 138 devices is assigned a unique device identifier. Table \ref{tbl:ModelsMultipleDevices} provides information on which models have more than one device represented in the WIDEFT corpus\footnote{In some cases of earbud pairs, the left earbud has a different model number (e.g.~A2031) than the right earbud (e.g.~A2032). For our purposes, we consider these two devices as having the same model (e.g.~A2031/A2032).}. Because some models/devices are capable of more than one wireless type and/or radio modes, 147 data captures were made. Each capture contains a set of 100 bursts. An index file \cite{zenodo2020} accompanies the WIDEFT corpus and provides additional device details such as make/manufacturer, model, center frequency, etc. 

\begin{table}[!htb]
    \centering
    \caption{WIDEFT Wireless Type Distribution. For additional device information such as make/manufacturer, model, center frequency, etc the reader is referred to the accompanying index file in the corpus.}
    \begin{adjustbox}{max width=0.99\linewidth,center}
    \begin{tabular}{lcc}
    \toprule
        \textbf{Wireless Type} & \textbf{Num.~Unique} & \textbf{Num.~Unique} \\
        \textbf{} & \textbf{Models} & \textbf{Device IDs} \\
            \midrule
        Bluetooth (2.4 GHz) & 64 & 81\\
        WiFi (2.4 GHz, 5 GHz) & 49 & 55\\
        Other (900 MHz) & 1 & 2\\
        \textbf{Total}$^1$          & \textbf{79} & \textbf{138}\\
    \bottomrule
    \end{tabular}
    \end{adjustbox}
    \label{tbl:WirelessTypeDistribution}
\end{table}

\begin{table}[!tbp]
    \centering
    \caption{Models in WIDEFT where data is collected from more than one device. Models in WIDEFT not listed below have data collected from only one device.}
     \begin{adjustbox}{max width=0.99\linewidth,center}
    \begin{tabular}{llc}
    \toprule
        \textbf{Wireless Type} & \textbf{Mfr.~Model} & \textbf{Num.~Devices}\\
            \midrule
        Bluetooth (2.4 GHz)& Apple A2031/A2032 & 8\\ 
        ~ & Aimson Q88 & 3\\
        ~ & Apple A1367 & 3\\
        ~ & Apple A2084/A2083 & 2\\ 
        ~ & Beats A2047/A2048 & 2\\ 
        ~ & Dell PP32LB & 2\\
        ~ & Haylou GT1 & 2\\
        ~ & Lenovo 81QB & 2\\
        ~ & MPOW BH437A & 2\\
        WiFi (2.4 / 5 GHz) & Aimson Q88 & 3\\
        ~ & Apple A1367 & 3\\
        ~ & Dell PP32LB & 2\\
        ~ & Lenovo 81QB & 2\\
        Other & iClicker RLR14 & 2\\
    \bottomrule
    \end{tabular}
    \end{adjustbox}
    \label{tbl:ModelsMultipleDevices}
\end{table}


%

\subsection{Data Collection Environment}
\label{ssec:DataCollection}

\begin{figure}[!t]
\centering
    \includegraphics[angle = 0,width = 0.9\linewidth]{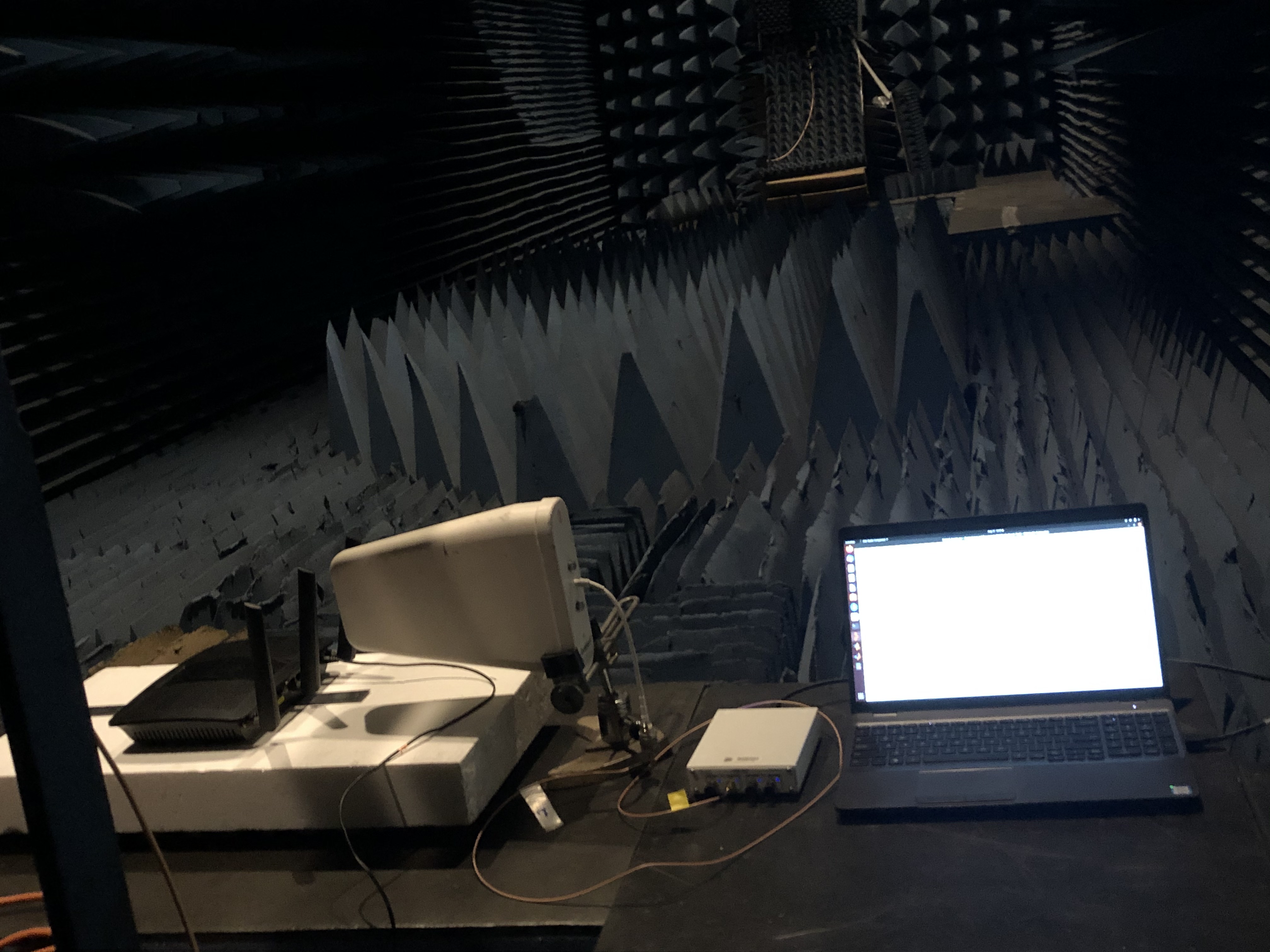}
    \caption{Photograph of the data acquisition setup consisting of a software defined radio (SDR) model USRP B210 connected to a PC using GNU Radio for radio device control and data acquisition.}    
    \label{fig:datacollectionenvironment}
\end{figure}

Our data acquisition system consisted of the USRP B210 SDR \cite{usrp2020}, a laptop computer, an antenna at the operating frequency of the device, and the device being examined. The laptop and SDR were connected via USB with the antenna connected to the SDR via a coax cable connected to the RX port. A stand was often used to hold the antenna steady to prevent any movement in the antenna during collection and to keep the collections as consistent as possible. To achieve the best SNR possible the bursts were collected inside an anechoic chamber located on the New Mexico State University campus. Our collection environment and data acquisition system, shown in Fig.~\ref{fig:datacollectionenvironment}, is similar to that used in \cite{uzundurukan2020}. For each device, we determined an orientation of the antenna which resulted in the highest signal power. A sampling rate of 56 MHz was used by the data acquisition system and the raw data recording was saved on a 3 GB RAM disk to ensure that the data could be saved at the fastest speed possible.

\section{Postprocessing and Data Organization}
\label{sec:postprocessingorganization}

\subsection{Signal Post-Processing}
\label{ssec:signalpostprocessing}
\begin{figure}[!t]
\centering
    \includegraphics[angle = 0,width = 0.9\linewidth]{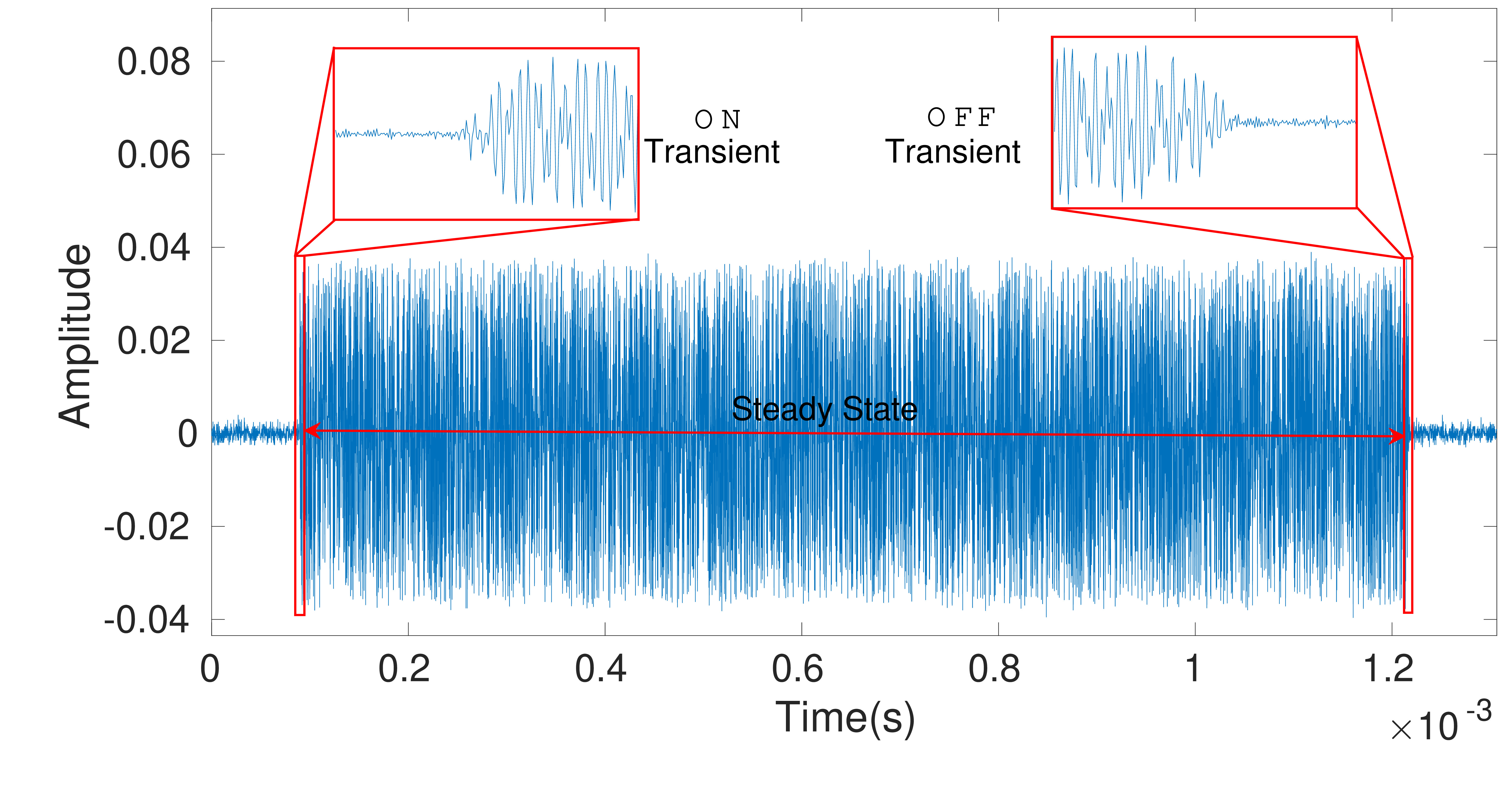}
    \caption{Plot of the signal burst with \on on and \off transients zoomed in for more detail.  This figure captures the period before (prefix) and after (suffix) the signal burst, the \on and  \off transients, and the steady-state portion of the signal.}
    \label{fig:waveformplot}
\end{figure}

In order to segment bursts from the raw recordings, we developed a {\sc Matlab} script to automate the burst detection. This script applied changepoint detection to the power spectrum to determine the beginning and end of the each radio burst. The signal segments provided in WIDEFT include 5000 samples before and after each burst. Fig.~\ref{fig:waveformplot} shows an example signal segment with the \on$\!\!$/\off transients and steady state, as well as 5000 samples before and after the transients. The signal bursts are saved as 16 bit I-Q sample pairs in \texttt{.sc16} file format. This process is repeated for each device until 100 distinct bursts have been collected and processed.

\subsection{Data Organization}
\label{ssec:DataOrganization}
The data files in WIDEFT are organized into a directory tree with the following path: \\ \ttcaps{WIDEFT/<WIRELESS\_TYPE>/<FREQUENCY>/}\\
\ttcaps{<PROTOCOL>/<DEVICE\_MAKE>/<DEVICE\_MODEL>\_}\\
\ttcaps{<DEVICE\_NUMBER>/<BURST\_ID>.sc16} where \\
\ttcaps{WIRELESS\_}
\ttcaps{TYPE} is `Bluetooth', `WiFi', or `Other';
\ttcaps{FREQUENCY} (WiFi only) is `2.4GHz' or `5GHz';
\ttcaps{PROTOCOL} (WiFi only) is '802.11a', '802.11b', '802.11g', '802.11n', '802.11ac', or 'Auto';
\ttcaps{DEVICE\_}
\ttcaps{MAKE} is the manufacturer of the device; 
\ttcaps{DEVICE\_MODEL} is the model number associated with the device;
\ttcaps{DEVICE\_}
\ttcaps{NUMBER} is used to differentiate multiple devices with the same model family; and 
\ttcaps{BURST\_ID} is an integer from 1 to 100.


The following are some examples of file paths and names for a single file.

\textbf{Bluetooth example:}

\noindent\ttcaps{WIDEFT/Bluetooth/Apple/A1296\_1/1.sc16}
translates to \ttcaps{Bluetooth} wireless type, manufactured by \ttcaps{Apple}, model number \ttcaps{A1296}, device \ttcaps{1}, burst ID \ttcaps{1}, \ttcaps{sc16} file.

\textbf{WiFi example:}

\noindent\ttcaps{WIDEFT/WiFi/2.4GHz/Auto/Dell/M4800\_1/\\ 15.sc16} translates to \ttcaps{WiFi} wireless type, \ttcaps{2.4GHz} frequency, \ttcaps{Auto} protocol, manufactured by \ttcaps{Dell}, model number \ttcaps{M4800}, device \ttcaps{1}, burst ID \ttcaps{15}, \ttcaps{sc16} file.

\textbf{Other example:}	

\noindent\ttcaps{WIDEFT/Other/iClicker/RLR14\_2/50.sc16} translates to \ttcaps{Other} wireless type, manufactured by \ttcaps{iClicker}, model number \ttcaps{RLR14}, device \ttcaps{2}, burst ID \ttcaps{50}, \ttcaps{sc16} file.

The \texttt{index.csv} file contains the associated information about the devices found in the corpus. N/A is used in the wireless protocol column for Bluetooth and Other devices. A `*' is used as a placeholder for the different burst number in the file name and path. The index file is broken up into the following columns in the given order: \ttcaps{DEVICE\_ID}, \ttcaps{PATH}, \ttcaps{WIRELESS\_TYPE}, \ttcaps{CENTER\_FREQUENCY}, \ttcaps{PROTOCOL}, \ttcaps{DEVICE\_MAKE}, \ttcaps{DEVICE\_MODEL}, \ttcaps{FCC\_ID}, \ttcaps{IC}, and \ttcaps{NOTES}.

\section{Classification of Device Fingerprints}
\label{sec:Classification}
In this section, we provide a baseline for classifier results using the the WIDEFT corpus. This includes a description of the feature vector whose elements are statistics of the instantaneous amplitude (IA), instantaneous phase (IP), and instantaneous frequency (IF) sequences during the \on transient. In addition, we describe the classifier and the results of the evaluations. 

We consider four different evaluations. First, we consider two device identification evaluations: 1) device identification using all 138 devices\footnote{For WiFi devices which have multiple operating modes, we use Auto and 2.4 GHz.} and 2) device identification using all eight available devices from model Apple A2031/A2032. Second, we consider two model identification evaluations: 3) model identification using all available devices per model and 4) model identification using only one device per model. For the remainder of the paper, we will denote these four evaluations as Evaluation 1, $\cdots$, Evaluation 4.

\subsection{Feature Extraction}
\label{ssec:features}
Previous research has considered features based on signal amplitude statistics \cite{rasmussen2007}, higher order statistics for use with common digital modulation schemes \cite{yuan2013specific}, cycle-frequency domain profile (statistical) for use with orthogonal frequency division multiplexing signals and tested on an IEEE 802.11a/g WLAN device \cite{kim2008specific}, and normalized permutation entropy \cite{huang2016specific}. Other work uses time-frequency and time-scale methods. For example features are considered based on wavelet coefficients extracted from the transients \cite{hippenstiel1996wavelet}, empirical mode decomposition and Haar wavelet decompositions \cite{song2010method}, Hilbert-Huang Transform \cite{yuan2014specific,  zhang2015novel, zhang2016specific}, ambiguity function and Wigner distribution \cite{li2011quadratic}, and the intrinsic time-scale decomposition \cite{song2010specific}. 

\begin{figure}[!htb]
    \centering
    {\includegraphics[clip,trim=1 60 1 80,width=0.5\textwidth]{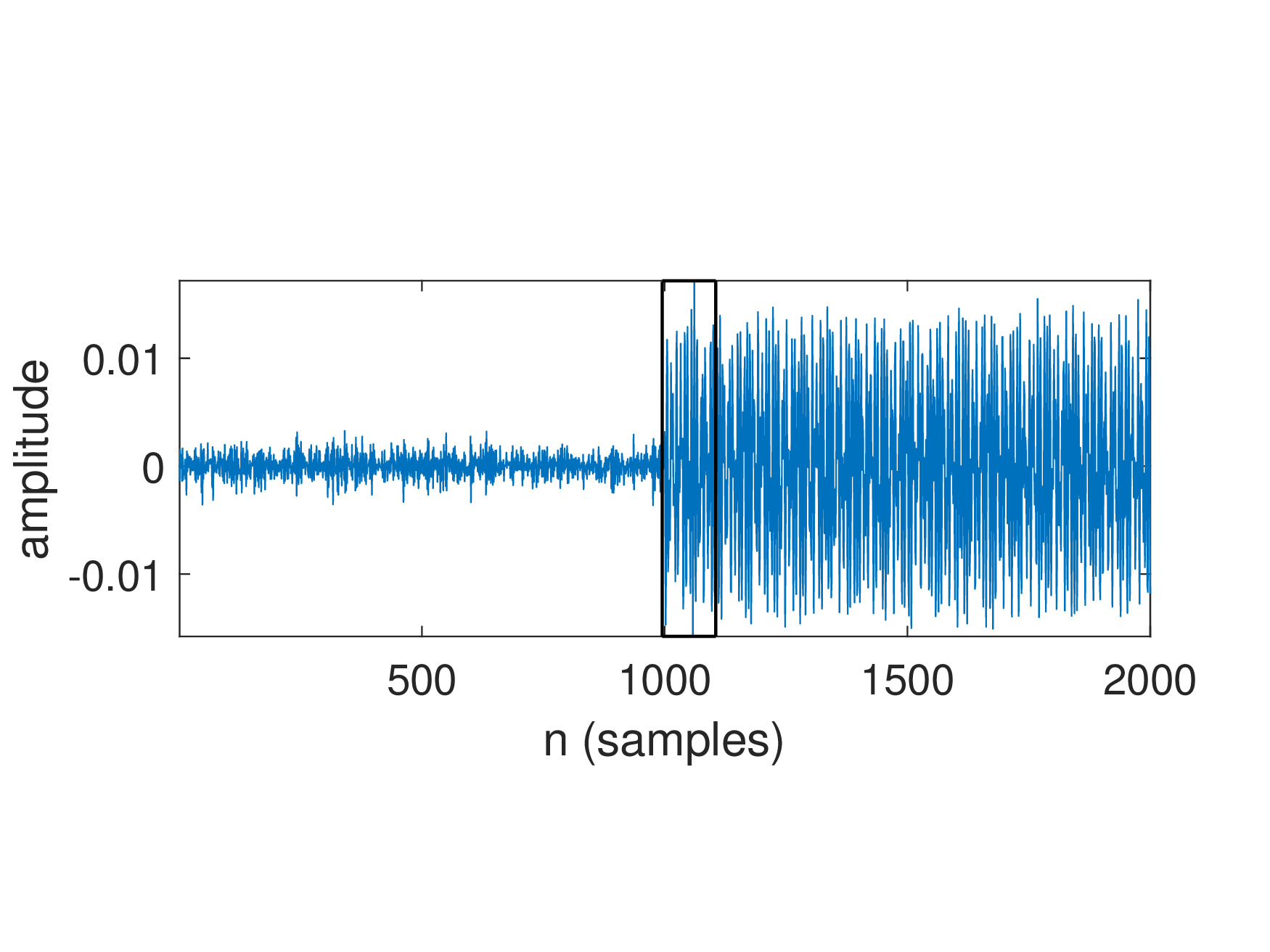}}
    \caption{An example of a single \on transient. The feature vector is computed using the region of the segment within the black frame.}
   \label{fig:tran}
\end{figure}

For each signal burst we detect the \on transient and segment the burst as follows. The analysis segment begins at the start of the \on transient and consists of 100 samples. This is illustrated in the black frame in Fig.~\ref{fig:tran}. For each segment we compute the corresponding IA, IP, and IF sequences \cite{sandoval2018instantaneous}. We compute statistics i.e.~mean, variance, skewness, and kurtosis from each instantaneous sequence (IA, IP, IF) and these form the elements of the 12-D feature vector \cite{wang2019identification, aghnaiya2019variational}.

\subsection{Classifier}
\label{ssec:classifier}
For each baseline evaluation, the data was randomly partitioned into two subsets, 75\% for training and 25\% for testing. We use an ensemble of 1000 tree classifiers with bagging, implemented using the \texttt{TreeBagger} function in {\sc Matlab}. 

\subsection{Evaluations and Results}
\label{ssec:results}
Table \ref{tbl:summary} shows the classification results for the four evaluation cases described earlier in this section. Each of these results provides an initial or baseline classification accuracy for the WIDEFT corpus for these evaluations. We have chosen a minimal subset of previously-proposed features and a simple classifier for these baseline evaluations \cite{wang2019identification, aghnaiya2019variational}. For Evaluation 1, where device identification was performed on all 138 devices, classification accuracy is 54.3\%. For Evaluation 2, where device identification was performed on eight Apple A2031/A2032 devices, classification accuracy is 82.0\% and the confusion matrix is given in Table \ref{tbl:confusion}. For Evaluation 3, where model identification was performed using all available devices per model, classification accuracy is 56.3\%. For Evaluation 4, model identification using only one device per model, classification accuracy is 62.7\%.

\begin{table}[!htb]
\centering
\renewcommand{\arraystretch}{1}
 \begin{adjustbox}{max width=1\linewidth,center}
\begin{tabular}{ccl}
\toprule
\textbf{Evaluation} & \textbf{Accuracy (\%)} & \textbf{Description}\\
\midrule
 1 &  54.3 & Device ID (all devices)\\ 
 2 &  82.0 & Device ID (Apple A2031/A2032)\\ 
 3 &  56.3 & Model ID (all devices)\\ 
 4 &  62.7 & Model ID (one device per model)\\ 
\bottomrule
\end{tabular}
\end{adjustbox}
\caption{Classification results for the four evaluations described in Section \ref{sec:Classification}.} 
\label{tbl:summary}
\end{table}

\begin{table}[!htb]
\centering
\renewcommand{\arraystretch}{1}
 \begin{adjustbox}{max width=0.99\linewidth,center}
\begin{tabular}{cp{4.45mm}p{4.45mm}p{4.45mm}p{4.45mm}p{4.45mm}p{4.45mm}p{4.45mm}p{4.45mm}}
\toprule
 & \multicolumn{8}{c}{\textbf{Predicted}}\\
  \cmidrule(lr){2-9}
\textbf{Actual} & \rotatebox{75}{A2031\_1} & \rotatebox{75}{A2031\_2} & \rotatebox{75}{A2031\_3} & \rotatebox{75}{A2031\_4} &  \rotatebox{75}{A2032\_1} & \rotatebox{75}{A2032\_2} & \rotatebox{75}{A2032\_3} & \rotatebox{75}{A2032\_4} \\ 
\midrule
 A2031\_1 &  \textbf{20} & ~1  &  ~$\cdot$ &  ~\,$\cdot$ & ~\,$\cdot$  &  ~4 &  ~\,$\cdot$ &  ~\,$\cdot$ \\ 
 A2031\_2 & ~1  &  \textbf{17} & ~\,$\cdot$ & ~\,$\cdot$  &  ~4 & ~\,$\cdot$  &  ~3 & ~\,$\cdot$  \\ 
 A2031\_3 &  ~\,$\cdot$ &  ~\,$\cdot$ &  \textbf{25} &  ~\,$\cdot$ & ~\,$\cdot$  &  ~\,$\cdot$ &  ~\,$\cdot$ &  ~\,$\cdot$ \\ 
 A2031\_4 & ~\,$\cdot$  &  ~\,$\cdot$ &  ~\,$\cdot$ &  \textbf{24} & ~\,$\cdot$  &  ~\,$\cdot$ &  ~\,$\cdot$ &  ~1 \\ 
 A2032\_1 &  ~1 &  ~4 & ~\,$\cdot$  & ~\,$\cdot$  &  \textbf{19} &  ~\,$\cdot$ &  ~1 &  ~\,$\cdot$ \\ 
 A2032\_2 &  ~8 & ~\,$\cdot$  & ~\,$\cdot$  & ~1  & ~\,$\cdot$  &  \textbf{16} &  ~\,$\cdot$ &  ~\,$\cdot$ \\ 
 A2032\_3 & ~1  &  ~4 &  ~\,$\cdot$ & ~\,$\cdot$ &  ~1 &  ~1 &  \textbf{18} &  ~\,$\cdot$ \\ 
 A2032\_4 &  ~\,$\cdot$ &  ~\,$\cdot$ & ~\,$\cdot$  &  ~\,$\cdot$ &  ~\,$\cdot$ &  ~\,$\cdot$ &  ~\,$\cdot$ &  \textbf{25} \\ 
\bottomrule
\end{tabular}
\end{adjustbox}
\caption{Confusion matrix for Evaluation 2 device identification (Apple A2031/A2032) described in Section \ref{sec:Classification}.} 
\label{tbl:confusion}
\end{table}

\section{Discussion}
\label{sec:discussion}
In this work, we have introduced the WIDEFT corpus with the intent of advancing research in the area of wireless device fingerprinting and specific emitter identification. The corpus consists of signal captures from a 138 wireless devices with each capture consisting of 100 wireless bursts. Unlike the recent corpus developed in \cite{uzundurukan2020} which includes only the \on transient and a portion of the steady state, the WIDEFT corpus consists of entire bursts, including the \on transient, steady state, and \off transient. The complete burst may allow for innovations in feature vector or classifier design for this research area. Additionally, while the corpus in \cite{uzundurukan2020} contains only Bluetooth smartphone devices, the WIDEFT corpus includes both Bluetooth and WiFi devices including smartphones, wireless speakers, headphones, earbuds, routers, keyboards, mice, and laptops. Finally, while the corpus in \cite{uzundurukan2020} contains signal captures from 27 unique smartphone models, the WIDEFT corpus contains signal captures from 79 unique models, including not just smartphones but other Bluetooth- / WiFi-enabled devices. This greater variety of device types and models allows a larger sample size when performing classifier evaluations and thus greater statistical power. On the other hand, increasing the number and variety of devices makes the problem more representative of the true ``population'' of wireless devices in current use. 

Evaluation 1 shows that a feature vector formed from statistics of the IA, IP, and IF sequences during the \on transient can be used to identify more than half (54.3\%) of the 138 unique devices. This experiment suggests that the fingerprint does have discriminating traits. Evaluation 2 considered a particular model, Apple A2031/A2032 which has the most example devices (eight) and shows that we can identify with 82.0\% accuracy. This experiment suggests that the fingerprint can identify devices within the same model family. An inspection of the confusion matrix in Table \ref{tbl:confusion} reveals that the false positives and false negatives for the first two devices (A2031\_1, A2031\_2) account for the majority of the errors. Evaluation 3 considered model identification using all devices. Unfortunately, because some models consist of several example devices and other models have only one example device, it is difficult to draw further conclusions. On the other hand, Evaluation 4 considered a model identification using a single device per model and thus has a balanced dataset (as compared to Evaluation 3). Assuming that in the feature space, the inter-model distance is much less than the intra-model distance, this can account for the increase in accuracy as compared to Evaluation 1. In all four evaluations, accuracy is far better than random guessing, however, there is considerable room for improvement in identification accuracy.


\section{Conclusions}
\label{sec:conclusion}
In this paper, we have introduced the publicly-available WIDEFT data corpus for advancing research in the area of wireless device fingerprinting and specific emitter identification. The WIDEFT corpus contains signal bursts from 138 unique devices (100 bursts per device), including 900 MHz, Bluetooth, and WiFi devices from 79 unique models. Additionally, to give an initial baseline of classifier performance for both device identification and model identification tasks using the WIDEFT corpus, we performed four evaluations. In these evaluations, we used previously-proposed features based on statistics computed from the instantaneous amplitude (IA), instantaneous phase (IP), and instantaneous frequency (IF) sequences during the \on transient of the radio burst. These evaluations provide baseline results for future research in which more sophisticated features and classification schemes may be compared against.


\bibliography{main} 

\begin{thebibliography}{10}
\providecommand{\url}[1]{#1}
\csname url@samestyle\endcsname
\providecommand{\newblock}{\relax}
\providecommand{\bibinfo}[2]{#2}
\providecommand{\BIBentrySTDinterwordspacing}{\spaceskip=0pt\relax}
\providecommand{\BIBentryALTinterwordstretchfactor}{4}
\providecommand{\BIBentryALTinterwordspacing}{\spaceskip=\fontdimen2\font plus
\BIBentryALTinterwordstretchfactor\fontdimen3\font minus
  \fontdimen4\font\relax}
\providecommand{\BIBforeignlanguage}[2]{{%
\expandafter\ifx\csname l@#1\endcsname\relax
\typeout{** WARNING: IEEEtran.bst: No hyphenation pattern has been}%
\typeout{** loaded for the language `#1'. Using the pattern for}%
\typeout{** the default language instead.}%
\else
\language=\csname l@#1\endcsname
\fi
#2}}
\providecommand{\BIBdecl}{\relax}
\BIBdecl

\bibitem{zhang2003}
Y.~Zhang, W.~Lee, and Y.-A. Huang, ``{Intrusion Detection Techniques for Mobile
  Wireless Networks},'' \emph{Wireless Networks}, vol.~9, no.~5, pp. 545--556,
  Sep. 2003.

\bibitem{banerjee2011}
S.~Banerjee and V.~Brik, \emph{Encyclopedia of Cryptography and
  Security}.\hskip 1em plus 0.5em minus 0.4em\relax Boston, MA: Springer US,
  2011, pp. 1388--1390.

\bibitem{hall2003detection}
J.~Hall, M.~Barbeau, and E.~Kranakis, ``{Detection of Transient in Radio
  Frequency Fingerprinting using Signal Phase},'' \emph{Wireless and Optical
  Commun.}, pp. 13--18, 2003.

\bibitem{danev2012physical}
B.~Danev, D.~Zanetti, and S.~Capkun, ``{On Physical-Layer Identification of
  Wireless Devices},'' \emph{ACM Computing Surveys}, vol.~45, no.~1, p.~6,
  2012.

\bibitem{sieka2006using}
Sieka and Bartlomiej, ``{Using Radio Device Fingerprinting for the Detection of
  Impersonation and Sybil Attacks in Wireless Networks},'' \emph{{Lecture Notes
  in Computer Science}}, vol. 4357, p. 179, 2006.

\bibitem{rasmussen2007}
K.~B. Rasmussen and S.~Capkun, ``{Implications of Radio Fingerprinting on the
  Security of Sensor Networks},'' in \emph{Proc.~Int.~Conf.~Security and
  Privacy in Commun.~Networks}, 2007, pp. 331--340.

\bibitem{ellis2001characteristics}
K.~Ellis and N.~Serinken, ``{Characteristics of Radio Transmitter
  Fingerprints},'' \emph{Radio Science}, vol.~36, no.~4, pp. 585--597, 2001.

\bibitem{vo2016fingerprinting}
T.~D. Vo-Huu, T.~D. Vo-Huu, and G.~Noubir, ``{Fingerprinting Wi-Fi Devices
  using Software Defined Radios},'' in \emph{Proc.~ACM Conf.~on Security and
  Privacy in Wireless and Mobile Networks}, 2016, pp. 3--14.

\bibitem{uzundurukan2020}
E.~Uzundurukan, Y.~Dalveren, and A.~Kara, ``{A Database for the Radio Frequency
  Fingerprinting of Bluetooth Devices},'' \emph{Data}, vol.~5, Jun. 2020.

\bibitem{zenodo2020}
\BIBentryALTinterwordspacing
(2020). [Online]. Available: \url{https://zenodo.org/record/4110980}
\BIBentrySTDinterwordspacing

\bibitem{usrp2020}
\BIBentryALTinterwordspacing
(2020). [Online]. Available:
  \url{https://files.ettus.com/manual/page_usrp_b200.html}
\BIBentrySTDinterwordspacing

\bibitem{yuan2013specific}
Y.-J. Yuan, Z.~Huang, and Z.-C. Sha, ``{Specific Emitter Identification based
  on Transient Energy Trajectory},'' \emph{Progress in Electromagnetics
  Research}, vol.~44, pp. 67--82, 2013.

\bibitem{kim2008specific}
K.~Kim, C.~M. Spooner, I.~Akbar, and J.~H. Reed, ``{Specific Emitter
  Identification for Cognitive Radio with Application to IEEE 802.11},'' in
  \emph{IEEE Global Telecommun.~Conf.}, 2008, pp. 1--5.

\bibitem{huang2016specific}
G.~Huang, Y.~Yuan, X.~Wang, and Z.~Huang, ``{Specific Emitter Identification
  Based on Nonlinear Dynamical Characteristics},'' \emph{Canadian J.~Elect.~and
  Comp.~Eng.}, vol.~39, no.~1, pp. 34--41, 2016.

\bibitem{hippenstiel1996wavelet}
R.~Hippenstiel and Y.~Payal, ``{Wavelet Based Transmitter Identification},'' in
  \emph{IEEE Sym.~Sig.~Process.~and its Applications}, vol.~2, 1996, pp.
  740--742.

\bibitem{song2010method}
C.~Song, J.~Xu, and Y.~Zhan, ``{A Method for Specific Emitter Identification
  based on Empirical Mode Decomposition},'' in \emph{IEEE Int.~Conf.~Wireless
  Commun., Networking and Info.~Security}, 2010, pp. 54--57.

\bibitem{yuan2014specific}
Y.~Yuan, Z.~Huang, H.~Wu, and X.~Wang, ``{Specific Emitter Identification based
  on Hilbert-Huang Transform-Based Time-Frequency-Energy Distribution
  Features},'' \emph{IET Commun.}, vol.~8, no.~13, pp. 2404--2412, 2014.

\bibitem{zhang2015novel}
J.~Zhang, F.~Wang, Z.~Zhong, and O.~Dobre, ``{Novel Hilbert Spectrum-Based
  Specific Emitter Identification for Single-Hop and Relaying Scenarios},'' in
  \emph{IEEE Global Commun.~Conf.}, 2015, pp. 1--6.

\bibitem{zhang2016specific}
J.~Zhang, F.~Wang, O.~A. Dobre, and Z.~Zhong, ``{Specific Emitter
  Identification via Hilbert-Huang Transform in Single-Hop and Relaying
  Scenarios},'' \emph{IEEE Trans.~Info.~Forensics and Security}, vol.~11,
  no.~6, pp. 1192--1205, 2016.

\bibitem{li2011quadratic}
L.~Li, H.-B. Ji, and L.~Jiang, ``{Quadratic Time--Frequency Analysis and
  Sequential Recognition for Specific Emitter Identification},'' \emph{IET
  Sig.~Process.}, vol.~5, no.~6, pp. 568--574, 2011.

\bibitem{song2010specific}
C.~Song, Y.~Zhan, and L.~Guo, ``{Specific Emitter Identification Based on
  Intrinsic Time-Scale Decomposition},'' in \emph{IEEE Int.~Conf.~Wireless
  Commun.~Networking and Mobile Computing}, 2010, pp. 1--4.

\bibitem{sandoval2018instantaneous}
S.~Sandoval and P.~L. De~Leon, ``{The Instantaneous Spectrum: A General
  Framework for Time-Frequency Analysis},'' \emph{IEEE Transactions on Signal
  Processing}, vol.~66, no.~21, pp. 5679--5693, 2018.

\bibitem{wang2019identification}
X.~Wang, Y.~Zhang, H.~Zhang, X.~Wei, and G.~Wang, ``{Identification and
  Authentication for Wireless Transmission Security Based on RF-DNA
  Fingerprint},'' \emph{EURASIP Journal on Wireless Communications and
  Networking}, vol. 2019, no.~1, p. 230, 2019.

\bibitem{aghnaiya2019variational}
A.~Aghnaiya, A.~M. Ali, and A.~Kara, ``{Variational Mode Decomposition-Based
  Radio Frequency Fingerprinting of Bluetooth Devices},'' \emph{IEEE Access},
  vol.~7, pp. 144\,054--144\,058, 2019.

\end{thebibliography}
\bibliographystyle{IEEEtran}

\vfill

\noindent \textbf{Biography of Presenting Author}

\noindent \textit{Abu Bucker Siddik} is a PhD student at the Klipsch School of Electrical and Computer Engineering at New Mexico State University (NMSU). He received his BS degree in electrical and electronic engineering from Khulna University of Engineering and Technology (KUET), Bangladesh in 2015. His research interests include signal processing, time-frequency analysis, atmospheric turbulence, and machine learning.

\end{document}